\documentclass[a4paper]{jpconf}
\usepackage{graphicx}
\usepackage{epstopdf}
\usepackage{subfigure}
\usepackage{textcomp}

\def\bkR{{\rm I\kern-.17em R}}
\def\bkC{{\rm \kern.24em \vrule width.05em height1.4ex depth-.05ex \kern-.26em C}}
\def\to{\rightarrow}

\def\be{\beta}

\def\frac#1#2{{\textstyle{{#1}\over {#2}}}}
\def\laq{\raise 0.4 ex \hbox{$<$}\kern -0.8 em\lower 0.62 ex\hbox{$\sim$}}
\def\gaq{\raise 0.4 ex \hbox{$>$}\kern -0.7 em\lower 0.62 ex\hbox{$\sim$}}

\def\be{\begin{equation}}
\def\ee{\end{equation}}
\def\ba{\begin{eqnarray}}
\def\ea{\end{eqnarray}}

\def\dalemb#1#2{{\vbox{\hrule height.#2pt
        \hbox{\vrule width.#2pt height#1pt \kern#1pt \vrule width.#2pt}
        \hrule height.#2pt}}}

\def\dalemb#1#2{{\vbox{\hrule height.#2pt
        \hbox{\vrule width.#2pt height#1pt \kern#1pt \vrule width.#2pt}
        \hrule height.#2pt}}}

\def\gtorder{\mathrel{\raise.3ex\hbox{$>$}\mkern-14mu
             \lower0.6ex\hbox{$\sim$}}}
\def\ltorder{\mathrel{\raise.3ex\hbox{$<$}\mkern-14mu
             \lower0.6ex\hbox{$\sim$}}}

\begin{document}

\title{\bf Noncommutative Black Holes \footnote{Based on a talk presented by CB at 1st Mediterranean Conference on Classical and Quantum Gravity 2009, Kolymabari, Crete, 14th - 18th September 2009, Greece.}}

\author{C Bastos$^{1,2}$, O Bertolami$^{1,2}$, N C Dias$^{3,4}$ and J N Prata$^{3,4} $}

\address{$^1$ Departamento de F\'\i sica, Instituto Superior T\'ecnico, Avenida Rovisco Pais 1, 1049-001 Lisboa, Portugal \\
  $^2$ Instituto de Plasmas e Fus\~ao Nuclear, Instituto Superior T\'ecnico, Avenida Rovisco Pais 1, 1049-001 Lisboa, Portugal \\
  $^3$ Departamento de Matem\'atica, Universidade Lus\'ofona de Humanidades e Tecnologias, Avenida Campo Grande, 376, 1749-024 Lisboa, Portugal \\
  $^4$ Grupo de F\'isica Matem\'atica, Universidade de Lisboa, Avenida Prof. Gama Pinto 2, 1649-003, Lisboa, Portugal}

\ead {cbastos@fisica.ist.utl.pt, orfeu@cosmos.ist.utl.pt, ncdias@mail.telepac.pt, joao.prata@mail.telepac.pt}

\begin{abstract}

{One considers phase-space noncommutativity in the context of a Kantowski-Sachs cosmological model to study the interior of a Schwarzschild black hole. It is shown that the potential function of the corresponding quantum cosmology problem has a local minimum. One deduces the thermodynamics and show that the Hawking temperature and entropy exhibit an explicit dependence on the momentum noncommutativity parameter, $\eta$. Furthermore, the $t=r=0$ singularity is analysed in the noncommutative regime and it is shown that the wave function vanishes in this limit.}

\end{abstract}

\section{Introduction}

The microscopic properties of black holes (BHs) and the singularity problem most likely require a quantum theory of gravity to be properly understood. Given that this theory is beyond reach, a quantum cosmology approach based on the minisuperspace approximation might be a helpful guideline. In this context noncommutative features should not be discarded. If so, one should seek for a noncommutative version of the Wheeler-DeWitt (WDW) equation for BHs. 

In this contribution one presents a study of the interior of a Schwarzschild BH \cite{Bastos5} using a phase-space noncommutativity of the Kantowski-Sachs (KS) cosmological model developed in Ref. \cite{Bastos2}. One obtains the temperature and the entropy of this noncommutative BH and examine the singularity $t=r=0$ in the context of this noncommutative model. 

A Schwarzschild BH is described by the metric
\be\label{eq0.1}
ds^2=-\left(1-{2M\over r}\right)dt^2+\left(1-{2M\over r}\right)^{-1}dr^2+r^2(d\theta^2+\sin^2\theta d\varphi^2)~,
\ee
where $r$ is the radial coordinate. For $r<2M$, within the horizon of events, the time and radial coordinates are interchanged ($r\leftrightarrow t$) and space-time is described by the metric
\be\label{eq0.2}
ds^2=-\left({2M\over t}-1\right)^{-1}dt^2+\left({2M\over t}-1\right)dr^2+t^2(d\theta^2+\sin^2\theta d\varphi^2)~.
\ee
That is, an isotropic metric turns into an anisotropic one, implying that the interior of a Schwarzschild BH can be described by an anisotropic cosmological space-time. Indeed, the metric (\ref{eq0.2}) can be mapped into the KS cosmological model \cite{AK}, which, in the Misner parametrization, can be written as
\be\label{eq1.4}
ds^2=-N^2dt^2+e^{2\sqrt{3}\beta}dr^2+e^{-2\sqrt{3}(\beta+\Omega)}(d\theta^2+\sin^2{\theta}d\varphi^2)~,
\ee
where $\Omega$ and $\beta$ are scale factors, and $N$ is the lapse function. The following identification for $t<2M$,
\be\label{eq0.3}
N^2=\left({2M\over t}-1\right)^{-1} \quad, \quad
e^{2\sqrt{3}\beta} = \left({2M\over t}-1\right) \quad , \quad
e^{-2\sqrt{3}\beta}e^{-2\sqrt{3}\Omega}=t^2~,
\ee
allows for mapping the metric Eq. (\ref{eq1.4}) into the metric Eq. (\ref{eq0.2}).

Thus, it is assumed that, at the quantum level, the interior of the BH can be described by the phase-space noncommutative  extension of the quantum KS cosmological model. Recently, the noncommutative Wheeler-DeWitt (NCWDW) equation of this problem was obtained and its solutions calculated explicitly. From this new NCWDW equation one computes the partition function for the Schwarzschild BH through the Feynman-Hibbs procedure \cite{Feynman}, in order to get the temperature and entropy of the BH. These quantities are explicitly dependent on the noncommuatative parameter, $\eta$ \cite{Bastos5}. Furthermore, the solutions of the NCWDW equation provide some insight about the BH singularity. It is shown that these solutions vanish in the neighbourhood of $t=r=0$, but that this does not necessarily imply a vanishing probability of finding the system at the singularity \cite{Bastos5}. One concludes that canonical phase-space noncommutativity is insufficient to ensure the avoidance of singularities. Interestingly, the obtained results suggest that the singularity may be removed through other forms of phase-space noncommutativity.

\section{Phase-Space Noncommutative Quantum Cosmology}

The ultimate structure of space-time should be determined by quantum gravity, and at this scale, presumably Planck scale, space-time might be noncommutative \cite{CS}. Before addressing the problem at hand, one discusses the mathematical background of the model.

Noncommutative quantum mechanics has been extensively discussed in the last few years \cite{BZAB}. A canonical extension of the Heisenberg-Weyl algebra is considered and time is assumed as being a commutative parameter. The theory is set in a $2d$-dimensional phase-space of operators with non-commuting position and momentum variables. Thus, the noncommutative algebra reads,\footnote{The units $c=\hbar=k_{B}=G=1$ are used.}
\be\label{eq1.1}
\left[\hat q_i, \hat q_j \right] = i\theta_{ij} \hspace{0.2 cm}, \hspace{0.2 cm} \left[\hat q_i, \hat p_j \right] = i  \delta_{ij} \hspace{0.2 cm},
\hspace{0.2 cm} \left[\hat p_i, \hat p_j \right] = i \eta_{ij} \hspace{0.2 cm},  \hspace{0.2 cm} i,j= 1, ... ,d
\ee
where $\eta_{ij}$ and $\theta_{ij}$ are antisymmetric real constant ($d \times d$) matrices and $\delta_{ij}$ is the identity matrix. The extended algebra is related to the standard Heisenberg-Weyl algebra:
\be\label{eq1.2}
\left[\hat R_i, \hat R_j \right] = 0 \hspace{0.2 cm}, \hspace{0.2 cm} \left[\hat R_i, \hat \Pi_j \right]
= i \hbar \delta_{ij} \hspace{0.2 cm}, \hspace{0.2 cm} \left[\hat \Pi_i, \hat \Pi_j \right] = 0 \hspace{0.2 cm},
\hspace{0.2 cm} i,j= 1, ... ,d ~,
\ee
by a class of linear (non-canonical) set of transformations:
\be\label{eq1.3}
\hat q_i = \hat q_i \left(\hat R_j , \hat \Pi_j \right) \hspace{0.2 cm},\hspace{0.2 cm}\hat p_i = \hat p_i \left(\hat R_j , \hat \Pi_j \right)
\ee
the so-called Seiberg-Witten (SW) maps\footnote{In the mathematics literature this mapping is usually referred to as Darboux map.}.

One reviews now some features of the phase-space noncommutative extension of the KS minisuperspace model \cite{Bastos2}. At least, two scale factors are needed in order to impose the noncommutative relations. Classically, one has the noncommutative Poisson algebra imposed on the scale factors $\beta$, $\Omega$ and on their conjugate momenta $P_{\beta}$, $P_{\Omega}$:
\ba\label{eq1.4.1}
\left\{\Omega, P_{\Omega} \right\} =1 \hspace{0.2 cm},\hspace{0.2 cm} \left\{\beta, P_{\beta} \right\} =1  \hspace{0.2 cm},\hspace{0.2 cm} \left\{\Omega, \beta \right\} = \theta  \hspace{0.2 cm},\hspace{0.2 cm} \left\{P_{\Omega}, P_{\beta} \right\} =\eta~.
\ea

Through the ADM formalism and taking $\Omega, \beta$ as configuration variables, one derives the Hamiltonian for this system,
\be\label{eq1.5}
H=N{\cal H}=Ne^{\sqrt{3}\beta+2\sqrt{3}\Omega}\left[-{P_{\Omega}^2\over24}+{P_{\beta}^2\over24}-2e^{-2\sqrt{3}\Omega}\right]~.
\ee
Choosing the lapse function associated to the KS metric, Eq. (\ref{eq1.4}), as $N=24e^{-\sqrt{3}\beta-2\sqrt{3}\Omega}$ one obtains a system of equations of motion for this problem \cite{Bastos2}. For the noncommutative Poisson algebra (\ref{eq1.4.1}), these read:
\ba\label{eq1.6}
&&\dot{\Omega}=-2P_{\Omega}~,\hspace{0.5cm}(a)\nonumber\\
&&\dot{P_{\Omega}}=2\eta P_{\beta}-96\sqrt{3}e^{-2\sqrt{3}\Omega}~,\hspace{0.5cm}(b)\nonumber\\
&&\dot{\beta}=2P_{\beta}-96\sqrt{3}\theta e^{-2\sqrt{3}\Omega}~,\hspace{0.5cm}(c)\nonumber\\
&&\dot{P_{\beta}}=2\eta P_{\Omega}~.\hspace{0.5cm}(d)
\ea
An analytical solution of this system is difficult to obtain given the entanglement of the variables. However, one can get a numerical solution that yields some predictions for the relevant physical quantities \cite{Bastos2}. Moreover, one finds that Eqs. (\ref{eq1.6}a) and (\ref{eq1.6}d) yield a constant of motion
\be\label{eq1.7}
\dot{P_{\beta}}=-\eta(-2P_{\Omega})=-\eta\dot{\Omega}\Rightarrow P_{\beta}+\eta\Omega=C~,
\ee
which is crucial to solve the phase-space NCWDW equation.

The canonical quantization of the classical Hamiltonian constraint, ${\cal H}\approx 0$, based on the ordinary Heisenberg-Weyl algebra, yields the commutative WDW equation for the wave function of the universe. For the simplest factor ordering of operators this equation reads
\be\label{eq1.8}
\left[- \hat P^2_{\Omega}+ \hat P^2_{\beta}-48e^{-2\sqrt{3} \hat{\Omega}}\right]\psi(\Omega,\beta)=0~.
\ee
where $\hat P_{\Omega}=-i \frac{\partial }{\partial \Omega}$, $\hat P_{\beta}=-i \frac{\partial }{\partial \beta}$ are the fundamental momentum operators conjugate to $\hat{\Omega} = \Omega$ and $\hat{\beta} = \beta$, respectively. The quantization of the classical algebra (\ref{eq1.4.1}), is achieved through the following noncommutative algebra:
\ba\label{eq1.8.1}
\left[\hat{\Omega}, \hat{P}_{\Omega} \right] =1  \hspace{0.2 cm},\hspace{0.2 cm} \left[\hat{\beta}, \hat{P}_{\beta} \right] =i \hspace{0.2 cm},\hspace{0.2 cm} \left[\hat{\Omega}, \hat{\beta} \right] = i \theta  \hspace{0.2 cm},\hspace{0.2 cm} \left[ \hat{P}_{\Omega}, \hat{P}_{\beta} \right] = i \eta~.
\ea

One can relate this noncommutative algebra with the Heisenberg-Weyl algebra via a SW map \cite{Bastos2}:
\ba\label{eq1.9}
\hat{\Omega} =\lambda \hat{\Omega}_{c}-{\theta\over2\lambda} \hat P_{\beta_c} \hspace{0.2cm} , \hspace{0.2cm} \hat{\beta} = \lambda \hat{\beta}_{c} + {\theta\over2\lambda} \hat P_{\Omega_c}~,\nonumber\\
\hat P_{\Omega}= \mu \hat P_{\Omega_c} + {\eta\over2\mu} \hat{\beta}_{c} \hspace{0.2cm} , \hspace{0.2cm} \hat P_{\beta}=\mu \hat P_{\beta_c}- {\eta\over2\mu} \hat{\Omega}_{c}~,
\ea
where the index $c$ denotes commutative variables, i.e. variables for which $\left[\hat{\Omega}_c, \hat{\beta}_c\right]=\left[\hat P_{\Omega_c}, \hat P_{\beta_c}\right]=0$ and $\left[\hat{\Omega}_c, \hat P_{\Omega_c}\right]=\left[\hat{\beta}_c, \hat P_{\beta_c}\right]=i$. It is possible invert the transformations Eqs. (\ref{eq1.9}) as long as
\be\label{eq1.91}
\xi \equiv \theta \eta <1.
\ee
The dimensionless constants $\lambda$ and $\mu$ satisfy the relationship \cite{Bastos2}
\be\label{eq1.10}
\left(\lambda\mu\right)^2-\lambda\mu+{\xi\over4}=0\Leftrightarrow\lambda\mu={1+\sqrt{1-\xi}\over2}~.
\ee
Thus, Eqs. (\ref{eq1.9}) allow for a representation of the operators (\ref{eq1.8.1}) as self-adjoint operators acting on the Hilbert space $L^2(\bkR^2)$. In this representation the WDW Eq. (\ref{eq1.8}) is deformed into a modified second order partial differential equation, which exhibits an explicit dependence on the noncommutative parameters:
\be\label{eq1.11}
\left[-\left(-i \mu {\partial \over \partial {\Omega_c}}+{\eta\over2\mu}\beta_{c}\right)^2+\left(-i \mu {\partial \over \partial {\beta_c}}-{\eta\over2\mu}\Omega_c\right)^2-48\exp{\left[-2\sqrt{3}\left(\lambda\Omega_c+i{\theta\over2\lambda} {\partial \over \partial {\beta_c}} \right)\right]}\right]\psi(\Omega_c,\beta_c)=0~.
\ee

From the constant of motion, Eq. (\ref{eq1.7}), one defines a new constant operator $\hat A=\frac{\hat C}{\sqrt{1-\xi}}$, from which follows that:
\be\label{eq1.12}
\mu \hat P_{\beta_c}+{\eta\over2\mu} \hat{\Omega}_c= \hat A~.
\ee
As this operator commutes with the noncommutative Hamiltonian of Eq. (\ref{eq1.11}), one looks for solutions of Eq. (\ref{eq1.11}) that are also eigenstates of $\hat A$. A generic eigenstate obeys the eigenvalue problem
\be\label{eq1.12a}
\left(-i\mu{\partial\over\partial\beta_c}+{\eta\over2\mu}\Omega_c\right)\psi_a(\Omega_c,\beta_c)=a \psi_a(\Omega_c,\beta_c)~.
\ee
where $a \in \bkR$ \cite{Bastos2}. Thus,
\be\label{eq1.12b}
\psi_a(\Omega_c,\beta_c)=\Re_a(\Omega_c)\exp{\left[{i\over\mu}\left(a-{\eta\over2\mu}\Omega_c\right)\beta_c\right]}~.
\ee
Substituting this solution into Eq. (\ref{eq1.11}) implies that $\phi_a(z)\equiv\Re_a(\Omega_c(z))$ satisfies:
\be\label{eq1.13}
\phi_a''(z)+\left(\eta z-a\right)^2\phi_a(z)-48\exp{[-2\sqrt{3}z+{\sqrt{3}\theta\over\lambda\mu}a]}\phi_a(z)=0~,
\ee
where one has introduced the new variable
\be\label{eq1.14}
z={\Omega_c\over\mu}\hspace{0.2 cm}\rightarrow\hspace{0.2 cm}{d\over dz}=\mu{d\over d\Omega_c}~.
\ee
So, one has found a second order ordinary differential equation which actually can be solved numerically. This equation depends on the eigenvalue $a$ and on the noncommutative parameters $\theta$ and $\eta$.

\section{Thermodynamics of Phase-Space Noncommutative Black Hole}

In order to get the temperature and the entropy for the phase-space noncommutative BH, one computes the partition function through the Feynman-Hibbs procedure \cite{Obregon}. This method is based on the minisuperspace potential function obtained from Eq. (\ref{eq1.13}):
\be\label{eq2.1}
V(z)=48\exp{[-2\sqrt{3}z+{\sqrt{3}\theta\over\lambda\mu}a]}-\left(\eta z-a\right)^2~.
\ee
After a change of variable, $x=z-{\theta\over2\lambda\mu}a$, the potential becomes
\be\label{eq2.3}
V(x)=48\exp{(-2\sqrt{3}x)}-\left(\eta x-c\right)^2~,
\ee
where $c=P_{\beta}(0)+\eta\Omega(0)$ is a constant from the classical constraint. 

%%%%%%%%%%%%%%%%%%%%%%%%%%%%%%%%%%%%%%
\begin{figure}
\begin{center}
\subfigure[ ~$\eta=0$ and $c=0.01$]{\includegraphics[scale=0.7]{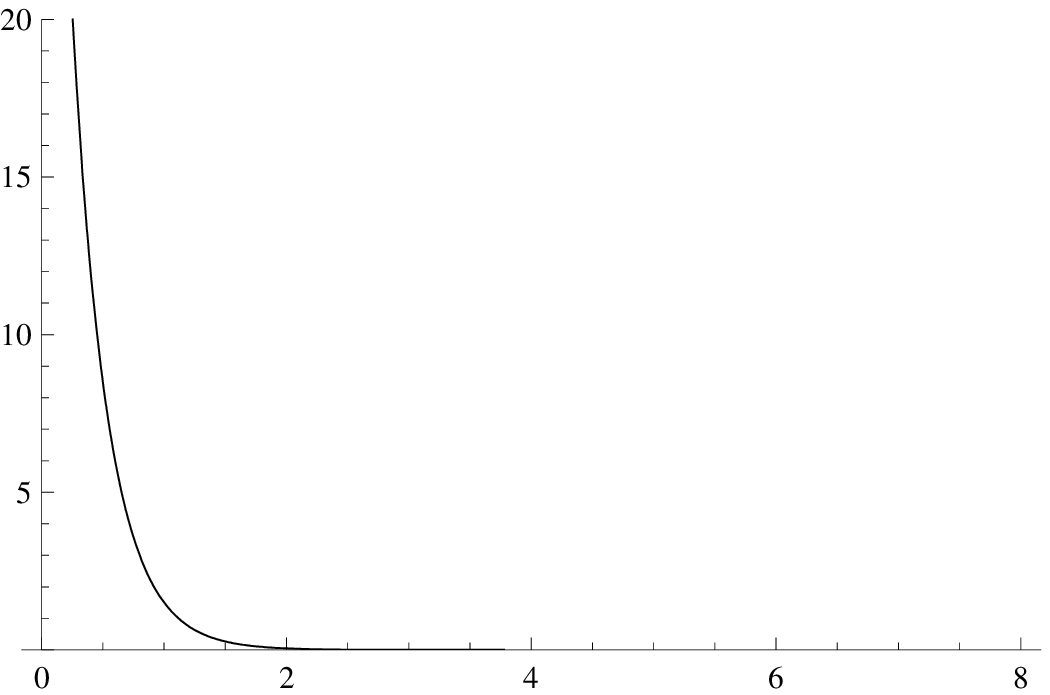}}
\subfigure[ ~$\eta=1.5$ and $c=5.68$]{\includegraphics[scale=0.7]{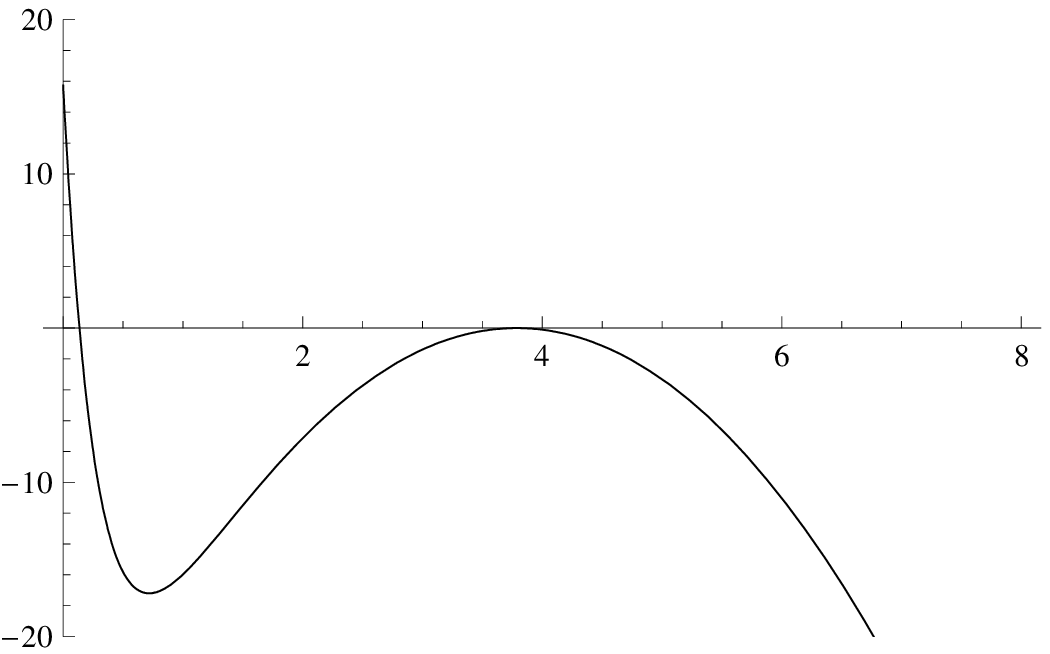}}
\caption{Potential function for some typical values of $\eta$ and $c$: (a) The potential function for the noncommutative case, $\theta\neq0$ and $\eta=0$ and (b) The potential function with $\eta\neq0$.  }
\label{fig:potential}
\end{center}
\end{figure}
%%%%%%%%%%%%%%%%%%%%%%%%%%%%%%%%%%%%%%

In Fig. \ref{fig:potential} one presents the potential function for the noncommutative case, $\theta\neq0$ and $\eta=0$, and the noncommutative case, where $\eta=1.5$. One sees that, qualitatively, there is no difference between the commutative case and the noncommutativity one for the configuration variables ($\eta =0$, $\theta \neq 0$) \cite{Dominguez}, that is, there is no local minimum. Thus, only when the noncommutativity in the momentum sector is introduced that one obtains a local minimum of the potential. The values of $\eta$ used in Fig. \ref{fig:potential} are fairly typical and the qualitative behaviour of the potential function is similar for other values. The constant $c$ is obtained using $P_{\beta}(0)=0.01$. Notice that the wave function, solution of the reduced NCWDW Eq. (\ref{eq1.13}), is well-defined for the chosen values of $\eta$ \cite{Bastos2}.

In order to use the Feynman-Hibbs method one expands the exponential term in the potential function in Eq. (\ref{eq2.3}) to second order in $(x-x_0)$, where the minimum $x_0$ is given by \cite{Bastos5}
\be\label{eq2.4}
{dV\over dx}|_{x_0}=-96\sqrt{3}\exp{(-2\sqrt{3}x_0)}-2 \eta (\eta x_0 - c)=0~.
\ee
The relationship,
\be\label{eq2.5}
\exp{(-2\sqrt{3}x_0)}=\zeta D-{\zeta^2\over\sqrt{3}}x_0~,
\ee
and the fact that the second derivative of the potential is positive, define the minimum of the potential function, 
\be\label{eq2.5a}
6\exp{(-2\sqrt{3}x_0)}-\zeta^2>0\Leftrightarrow x_0< -{1 \over \sqrt{3}} \ln \left({\zeta \over \sqrt{6}}\right) ~,
\ee
where $\zeta=\eta/4\sqrt{3}$ and $D=c/12$.
Thus, the NCWDW equation becomes \cite{Bastos5}
\be\label{eq2.7}
-{1\over2}{d^2\phi\over dx^2}+24(6e^{-2\sqrt{3}x_0}-\zeta^2)(x-x_0)^2\phi+[24 e^{-2\sqrt{3}x_0} - {1 \over 2}(\eta x_0-c)^2]\phi=0~.
\ee
This equation is analogous to the Schr\"odinger equation of the harmonic oscillator. So, one can identify the potential function as
\be\label{eq2.8}
V_{NC}(y)=24(6e^{-2\sqrt{3}x_0}-\zeta^2)y^2~,
\ee
where $y=x-x_0$. The Feynman-Hibbs procedure allows for introducing quantum corrections to the partition function through the potential, i.e. the quantum correction to the potential is given by \cite{Feynman},
\be\label{eq2.9}
{\beta_{BH}\over24}V_{NC}''(y)=2\beta_{BH}(6e^{-2\sqrt{3}x_0}-\zeta^2)~,
\ee
where $\beta_{BH}$ is the inverse of the BH temperature. The noncommutative potential used to compute the partition function is the following,
\be\label{eq2.10}
U_{NC}(y)=24(6e^{-2\sqrt{3}x_0}-\zeta^2)\left(y^2+{\beta_{BH}\over12}\right)~.
\ee

Thus, the noncommutative partition function is given by
\be\label{eq2.11}
Z_{NC}=\sqrt{1\over{48(6e^{-2\sqrt{3}x_0}-\zeta^2)}}{1\over\beta_{BH}}\exp{\left[-2{\beta_{BH}^2}\left(6e^{-2\sqrt{3}x_0}-\zeta^2\right)\right]}~.
\ee
The noncommutative internal energy of the BH, $\bar{E}_{NC}=-{\partial\over\partial\beta_{BH}}\ln Z_{NC}$, is given by
\be\label{eq2.12}
\bar{E}_{NC}={1\over\beta_{BH}}+4(6e^{-2\sqrt{3}x_0}-\zeta^2)\beta_{BH}~.
\ee
Setting that $\bar{E}_{NC}=M$, one derives the BH temperature
\be\label{eq2.13}
\beta_{BH}={M\over8(6e^{-2\sqrt{3}x_0}-\zeta^2)}\left[1 \pm \left(1 - {16\over M^2}(6e^{-2\sqrt{3}x_0}-\zeta^2)\right)^{1/2}\right]~.
\ee
The BH temperature is given by inverting, Eq. (\ref{eq2.13}). Assuming that $M>>1$ and taking the positive root, then:
\be\label{eq2.13a}
T_{BH}={4 \over M}(6e^{-2\sqrt{3}x_0}-{\zeta^2})~.
\ee
Notice that this quantity must be positive (cf. Eq. (\ref{eq2.5a})). To compare this result with the Hawking temperature $T_{BH}= {1 \over 8 \pi M}$, one should be cautious as the limit $\eta \rightarrow 0$ is ill-defined. Notice that Eq. (\ref{eq2.13a}) has the same mass dependence as the Hawking temperature 
\be\label{eq2.13.1a}
T_{BH}={b(\zeta)\over M}~,
\ee
where $b(\zeta) = 4 (6e^{-2\sqrt{3}x_0}-{\zeta^2})$.

One can recover the Hawking temperature even in the presence of the momentum noncommutativity for a specific value of $\eta$. Indeed, equating (\ref{eq2.13a}) with the Hawking temperature and using the stationarity condition Eq. (\ref{eq2.5}), one gets for $c=12 D =5.68$:
\be\label{eq2.13.1c}
x_0=1.8478 \hspace{1 cm} \eta=0.025~.
\ee
Since $\eta$ cannot be exactly equal to zero, one uses $\eta_0=0.025$ as a reference value which yields the Hawking temperature. As $\eta$ increases one gets a gradual noncommutative deformation of the Hawking temperature.

Finally, through the relationship, $S_{NC}=\ln Z_{NC}+\beta_{BH}\bar{E}_{NC}$ one obtains the phase-space noncommutative BH entropy: 
\ba\label{eq2.14}
S_{BH}&=&\ln{1\over\sqrt{12 b(\zeta)}}+{M^2\over{2 b(\zeta)}}\left(1+\sqrt{1-{4b(\zeta)\over M^2}}\right)-{M^2\over{8 b(\zeta)}}\left(1+\sqrt{1-{4 b(\zeta)\over M^2}}\right)^2+\nonumber\\
&-&\ln{M\over{2 b(\zeta)}}\left(1+\sqrt{1-{4 b(\zeta)\over M^2}}\right)~.
\ea
Once again, as $M>>1$, neglecting terms proportional to $M^{-2}$ yields
\be\label{eq2.16}
S_{BH}\simeq{M^2\over2 b(\zeta)}+\ln{\sqrt{b(\zeta)}\over{M \sqrt{3}}}~.
\ee
For $\eta=\eta_0$, $b(\zeta_0) = 1/( 8 \pi)$, one recovers the Hawking entropy and some "stringy" corrections
\be
S_{BH}=4\pi M^2+\ln{\sqrt{2 \pi\over3}}-\ln({8\pi M})~.
\ee\label{eq2.16a}

\section{Singularity}

One examines now the $r=t=0$ singularity. Notice that one employs the KS metric to describe the interior of the Schwarzschild BH through the identification Eq. (\ref{eq0.3}). Thus, $t=0$ corresponds to $\Omega\rightarrow +\infty$ and $\beta\rightarrow +\infty$. Therefore, one is interested in studying the limit
\be\label{eq3.1.1}
\lim_{\Omega_c,\beta_c \to +\infty} \psi(\Omega_c,\beta_c)~,
\ee
where $\psi(\Omega_c,\beta_c)$ is a generic solution of Eq. (\ref{eq1.11}). In terms of the eigenstates of $\hat A$ Eq. (\ref{eq1.12}), the solutions of Eq. (\ref{eq1.11}) can be represented by
\be\label{eq3.1.2}
\psi(\Omega_c,\beta_c)= \int da \, C(a) \psi_a(\Omega_c,\beta_c)
\ee
where $C(a)\in \bkC$ and $\psi_a(\Omega_c,\beta_c)$ is of the form
\be\label{eq3.1.3}
\psi_a(\Omega_c,\beta_c)=\phi_a \left({\Omega_c\over\mu}\right) \exp\left[{i\over\mu}\left(a-{\eta\over2\mu}\Omega_c \right) \beta_c \right]
\ee
and $\phi_a(z)$, $z=\Omega_c/\mu$ satisfies Eq. (\ref{eq1.13}). For $\Omega \rightarrow \infty$, one keeps only the dominant terms, such as
\be\label{eq3.2}
\phi_a''(z)+\left(\eta z-a\right)^2\phi_a(z)=0~.
\ee
This equation can be rewritten for $\eta\neq0$ as
\be\label{eq3.3}
\left\{-{\partial^2\over\partial \tilde z^2}-\eta^2 \tilde z^2 \right\} \tilde\phi_a(\tilde z)=0~,
\ee
where one has performed a change of variables, $\tilde z=z-\frac{a}{\eta}$ and $\tilde\phi_a(x)=\phi_a(x+\frac{a}{\eta})$. As can be clearly seen, this equation is similar to the eigenvalue equation of an inverted harmonic oscillator. This Hamiltonian is self-adjoint in $L^2(\bkR^2)$, its spectrum is continuous and its zero eigenfunction (the solution of Eq. (\ref{eq3.3})) displays the asymptotic form (for $\eta \not=0$) \cite{Gitman1} 
\be\label{eq3.4}
\tilde\phi_a(\tilde z)\sim {1\over\tilde z^{1/2}}\exp \left[\pm i {\eta\over2}\tilde z^2 \right]
\ee
and so, for all $a$,
\be\label{eq3.5}
\lim_{z \to +\infty} \phi_a(z) = \lim_{z \to +\infty} \tilde\phi_a(z-{a\over\eta})=0 \quad \Longrightarrow \quad \lim_{\Omega_c,\beta_c \to +\infty} \psi_a(\Omega_c,\beta_c)=0
\ee
Therefore, considering a fairly general choice of coefficients $C(a)$, it is expected that 
\be\label{eq3.6}
\lim_{\Omega_c,\beta_c \to +\infty} \psi(\Omega_c,\beta_c)=0~,
\ee
which is a necessary condition to provide a quantum regularization of the classical singularity of the Schwarzschild BH. However, one should be cautious before concluding that the probability of finding the BH at the singularity is zero. Although the calculation of probabilities for general covariant systems is a delicate issue, in here, given that the wave function is oscillatory in $\beta_c$, it is natural to fix a $\beta_c$-hypersurface, corresponding to the introduction of the measure $\delta(\beta-\beta_c) d \beta d\Omega_c$ in the probability distribution. The probability $P(r=0,t=0)$ of finding the BH at the singularity is then given by
\be\label{eq3.7}
P(r=0,t=0)= \lim_{\Omega_c,\beta_c \to +\infty} \int_{\Omega_c}^{+\infty}|\psi(\Omega_c',\beta_c)|^2 d \Omega_c'\simeq \lim_{\Omega_c \to +\infty}  \int_{\Omega_c}^{+\infty}|\phi_a({\Omega'_c\over\mu})|^2d\Omega_c'
\ee
which, unfortunately, is divergent. This follows from the conclusion (which can be derived from the asymptotic expression) that the inverted harmonic oscillator displays non-normalizable eigenstates. Hence, the noncommutativity of the form (\ref{eq1.8.1}) cannot be regarded as the final answer for the singularity problem of the Schwarzschild BH.

However, if one considers instead a Hamiltonian like
\be\label{eq3.8}
H= -{\partial^2\over\partial x^2} +V(x) \quad , \quad V(x) \sim -\eta^2 x^{2+2\epsilon}
\ee
for some $\epsilon >0$, one concludes that it displays zero energy eigenstates solutions of the \cite{Gitman}
\be\label{eq3.9}
\psi(x) \sim {1\over x^{(1+\epsilon)/2}}\exp \left[\pm i \frac{\eta}{2+\epsilon}\tilde x^{2+\epsilon} \right]
\ee
which are normalizable. One can then conjecture that a suitable deformation of the noncommutative structure (\ref{eq1.8.1}) may lead to a NCWDW equation associated to a potential of the form $V(x)\sim -\eta^2 x^{2(1+\epsilon)}$ for some arbitrarily small $\epsilon >0$. The solutions of this new NCWDW equation would then display vanishing probability (in the sense of Eq. (\ref{eq3.7})) at the singularity, thus solving this problem for the Schwarzschild BH \cite{Bastos6}.

\section{Conclusions}

In this contribution the temperature and the entropy of a phase-space noncommutative Schwarzschild BH have been computed. This has been performed through a KS cosmological model and the NCWDW equation of the problem.

The $t=r=0$ singularity was examined in the same context and found that for $t=0$, or $\Omega,\beta\rightarrow+\infty$, one encounters the Schr\"odinger problem of the inverted harmonic oscillator. The wave function vanishes in this limit, but given that it is not square integrable, it is not possible to conclude that the probability vanishes at the singularity. However, the discussed approach suggests that this feature may be achieved through more general noncommutativity relations \cite{Bastos6}.

\ack

\noindent The work of CB is supported by Funda\c{c}\~{a}o para a Ci\^{e}ncia e a Tecnologia (FCT) under the fellowship SFRH/BD/24058/2005. The work of NCD and JNP was partially supported by Grant No. POCTI/0208/2003 and PTDC/MAT/69635/2006 of the FCT.

%\vspace{0.3cm}

\end{document}